\documentclass[aps,superscriptaddress,eqsecnum,nofootinbib,showpacs,preprintnumbers]{revtex4-2}
\pdfoutput=1
\usepackage[utf8]{inputenc}
\usepackage{amsmath}
\usepackage{amssymb}
\usepackage{amsfonts,amsthm,bm}
\usepackage{color,xcolor}
\usepackage{comment}
\usepackage{soul}

\usepackage{graphicx,epsfig}
\usepackage{float}
\usepackage{subcaption}
\usepackage{tikz}
\newcommand{\be}{\begin{eqnarray}}
	\newcommand{\ee}{\end{eqnarray}}
\newcommand{\bea}{\begin{eqnarray}}
	
	\newcommand{\eea}{\end{eqnarray}}

\usepackage{ulem}
\definecolor{azure(colorwheel)}{rgb}{0.0, 0.5, 1.0}
\definecolor{DarkViolet}{RGB}{148,0,211}
\definecolor{myDarkBlue}{rgb}{0,0.1,0.7}
\definecolor{DarkBlue}{RGB}{0,0,153}
\definecolor{amber}{rgb}{1.0, 0.49, 0.0}
\definecolor{amaranth}{rgb}{0.9, 0.17, 0.31}
\definecolor{nicered}{rgb}{0.7,0.1,0.1}
\definecolor{brown}{rgb}{0.5,0.1,0.1}
\definecolor{nicegreen}{rgb}{0.0,0.3,0.0}
\definecolor{tealgreen}{rgb}{0.0, 0.51, 0.5}

\definecolor{tclr}{RGB}{148,0,211}
\usepackage{hyperref}
\hypersetup{colorlinks,bookmarksopen,
	bookmarksnumbered,
	citecolor={nicered},
	linkcolor={myDarkBlue},
	urlcolor={tealgreen},
	pdfstartview=FitH}

\newcommand{\beq}{\begin{equation}}
\newcommand{\eeq}{\end{equation}}
\newcommand{\bseq}{\begin{subequations}}
	\newcommand{\eseq}{\end{subequations}}

\usepackage{times}

\graphicspath{{Figs/}}

\usepackage{orcidlink}
\def\idbako{\orcidlink{0000-0002-3012-6144}}

\def\idkara{\orcidlink{0000-0002-5479-6513}}
\def\idpapa{\orcidlink{0000-0003-1244-922X}}
\pdfoutput=1

\begin{document}
\title{To stealth or not to stealth: Thermodynamics of stealth black holes}

\author{Athanasios Bakopoulos\idbako}
\email{atbakopoulos@gmail.com}
\affiliation{Physics Department, School of Applied mathematical and Physical Sciences,
	National Technical University of Athens, 15780 Zografou Campus,
	Athens, Greece.}

\author{Thanasis Karakasis\idkara}
	\email{thanasiskarakasis@mail.ntua.gr}
	\affiliation{Physics Department, School of Applied mathematical and Physical Sciences,
	National Technical University of Athens, 15780 Zografou Campus,
	Athens, Greece.}

\author{Eleftherios Papantonopoulos\idpapa}
	\email{lpapa@central.ntua.gr} 

\affiliation{Physics Department, School of Applied mathematical and Physical Sciences,
	National Technical University of Athens, 15780 Zografou Campus,
	Athens, Greece.}

\begin{abstract}
    Using Euclidean methods we investigate the thermodynamics of stealth black hole solutions, which are solutions that geometrically are indistinguishable from those of general relativity, however they are accompanied by a non-trivial additional field that does not back-react to the metric. We find that in general, one can observe shifts in the mass and/or the entropy of such black holes, hence, at least at the thermodynamic level, the stealth solutions may be distinguished by those of general relativity. We also point out an example, the {\it{bona fide stealth}}, where a non-trivial scalar field can accompany an (A)dS-Schwarzschild black hole without altering the thermodynamic quantities. We point out that thermodynamic discrepancies arise due to the additional boundary terms emanating from the stealth fields, where care should be given in order for the theory at hand to possess a well-defined variational procedure. 
\end{abstract}

\maketitle

\section{Introduction}

Black holes have long served as an invaluable testing ground for both general relativity and its possible extensions. As astrophysical objects that harbor extreme gravitational fields, they provide a unique setting for probing the fundamental nature of gravity. While general relativity has stood up to numerous observational tests, the quest for a quantum theory of gravity and the growing interest in dark energy and dark matter have driven the exploration of modified theories of gravity. Among these, scalar-tensor theories have emerged as prominent candidates, where gravity is mediated not only by the spacetime metric but also by a scalar field. These theories provide a natural extension of Einstein's gravity, introducing new degrees of freedom that could manifest in strong gravitational regimes, such as near black holes.

One of the more intriguing discoveries within this framework is the existence of stealth black holes. These are black hole solutions where, despite the presence of a non-trivial scalar field, the external spacetime geometry remains indistinguishable from that of a black hole in general relativity. Specifically, in scalar-tensor theories such as Horndeski and its generalizations, stealth solutions have been found where the scalar field coexists with the well-known Schwarzschild or Kerr geometries, yet without altering the observable properties of the black hole. Stealth solutions have also been found in theories with gauge fields. As a result, stealth black holes evade detection by traditional observational means, effectively ``hiding" the deviations from general relativity.

In particular, considering a gravity theory with additional fields (for example scalar or gauge fields) and a cosmological constant, one can write the Einstein equations in the Einstein frame as
\begin{equation}
    G_{\mu\nu} + \Lambda g_{\mu\nu} = T_{\mu\nu}~, \label{EE}
\end{equation}
where $T_{\mu\nu}$ denotes the energy-momentum tensor that arises when varying the extra fields with respect to the metric. A stealth solution is therefore one which realises 
\begin{equation}
    G_{\mu\nu} + \Lambda g_{\mu\nu} = 0 \hspace{0.5cm} \text{and} \hspace{0.5cm} T_{\mu\nu} = 0~, \label{stealthcondition}
\end{equation}
and hence the result is an Einstein manifold, accompanied by additional fields that do not affect the spacetime metric. 
The first stealth solution was reported by Ayon-Beato, Martinez and Zanelli \cite{Ayon-Beato:2004nzi} where a time dependent non-minimally coupled scalar field was found to overfly the non-rotating BTZ black hole \cite{Banados:1992wn}. In their theory the authors, considered a scalar field, with its kinetic term and potential, with an additional coupling between the scalar field and curvature of the form $\xi R\phi^2$. This particular coupling results in the emanation of second order differential equations for the scalar field. The $\xi R\phi^2$ coupling, has been shown to also give rise to stealth solutions in three-dimensional massive gravity \cite{Hassaine:2013cma}, as well as $1+1$ dilaton gravity \cite{Alvarez:2016qky}. The scenario of stealth fields was then reconsidered in the case of flat spacetime \cite{Ayon-Beato:2005yoq} and it is extended in \cite{Ayon-Beato:2024xgp} in order to include the most general conformal contributions that can be considered, the AdS-black-hole case was tackled in \cite{Caldarelli:2013gqa}, while planar stealth solutions were also found in \cite{BravoGaete:2013acu}. Very recently, it was shown that the $R\phi^2$ coupling can support stealth black holes in a plethora of theories \cite{Erices:2024iah}. In 2013, the first stealth black hole in shift and parity symmetric Horndeski theory was obtained, considering a time-dependent scalar field \cite{Babichev:2013cya,Kobayashi:2014eva}. The authors considered a part of Horndeski theory that possesses shift and parity symmetry and obtained the Schwarzschild and AdS Schwarzschild black holes with a non-trivial scalar field in the background that does not back-react to the spacetime element. The shift and parity symmetry allow for a linear dependent scalar field in the form of $\phi(t,r) = q t + \psi (r)$ where $q$ is an integration constant arising from the requirement of shift symmetry, preserving the staticity of the spacetime metric, as well as the finiteness of the action of the theory for the static black holes it hosts. For the three-dimensional version of this case we refer to \cite{Bravo-Gaete:2014haa}, while several stealth solutions have been obtained in Horndeski theories in various scenarios \cite{Minamitsuji:2018vuw, Minamitsuji:2019shy, Bernardo:2019yxp, Bakopoulos:2023fmv, Baake:2023zsq, Bakopoulos:2023sdm}, with their perturbations been studied in \cite{deRham:2019gha, Bernardo:2020ehy}, where strong-coupling problems were encountered. 

Stealth black hole solutions have also been obtained in the class of theories denoted as vector-tensor theories \cite{Chagoya:2016aar,Heisenberg:2017xda}, that were proposed to explain dark energy by breaking the Abelian symmetry, by coupling vector fields to gravity non-minimally \cite{Gripaios:2004ms, Tasinato:2014eka}. In simple cases of these theories, it is possible to find a stealth Schwarzchild black hole solution \cite{Heisenberg:2017hwb, Chagoya:2017ojn, Minamitsuji:2016ydr}, with a non-trivial gauge field. In particular, such a solution arises from the non-minimal coupling $\sim G_{\mu\nu}A^{\mu}A^{\nu}$ where $A^{\mu}$ is the gauge field, which is dynamical, but does not affect the spacetime metric. It is clear that introducing such a term in the action breaks the gauge symmetry of $U(1)$, since now the Lagrangian will not be invariant under the transformation $A_{\mu} \to A_{\mu} + \partial_\mu \Omega(x)$. This particular scenario will be a central part of this work. Finally, in the context of cosmology, stealth configurations have been found to be of particular interest. In \cite{Ayon-Beato:2013bsa, Ayon-Beato:2015mxf} it is pointed out that any standard FRW cosmology can be accompanied by a non-minimally coupled scalar field, without causing any back-reaction on spacetime. The creation probability of a de-Sitter universe with a stealth scalar field was investigated in \cite{Maeda:2012tu}, while it has been also shown that the $\Lambda-$ Cold Dark Matter model ($\Lambda$CDM) can be mimicked by a stealth scalar field. However, in this last case, we do not have the standard stealth scenario, because here the stealth scalar field plays the role of the matter content of the universe.

Stealth black holes introduce a novel concept within the landscape of modified gravity theories. In these scenarios, the extra field (scalar or gauge), though present, has no observable impact on the black hole’s external geometry—it leaves the geometry unaffected, and hence what can be deduced by the geometry itself, will also remain unaffected. For an external observer, the black hole appears indistinguishable from its counterpart in general relativity. The physical significance of stealth black holes lies in their challenge to our understanding of gravity. These objects demonstrate that certain modified gravity theories can allow the existence of non-trivial fields, yet the usual observational signals associated with black holes remain the same as in general relativity. This highlights a potentially profound limitation in our ability to probe the nature of gravity through black hole observations.

Stealth black holes are particularly important in the context of testing gravity. If modified gravity theories can hide their effects in this way, it raises the possibility that deviations from general relativity may be more subtle than expected, potentially limiting the sensitivity of current observational tests. Traditional methods of testing Einstein's theory rely on the detection of deviations in the behavior of objects in extreme gravitational environments. However, the existence of stealth black holes suggests that certain modifications to gravity might not manifest in any observable way within black hole spacetimes. This introduces a significant obstacle in the quest to constrain or falsify alternative theories of gravity. While stealth black holes may appear indistinguishable from Schwarzschild or Kerr black holes in terms of their gravitational effects, they could still produce observable signatures indirectly. For instance, black hole mergers may involve scalar fields becoming dynamical, potentially leaving detectable imprints on gravitational waves. In the same manner cosmological perturbations could reveal deviations from general relativity on larger scales. Thus, although stealth black holes are invisible to traditional methods, they remain a crucial target in the broader effort to test the limits of Einstein's gravity. %
In cosmology, stealth black holes also play an intriguing role. Scalar-tensor theories often feature prominently in models that attempt to explain cosmic acceleration, dark energy, and other large-scale phenomena. The presence of stealth black holes in these theories suggests that black hole solutions in modified gravity can exist without disturbing the large-scale structure of spacetime. This could have implications for the evolution of the universe, particularly in scenarios where black hole formation and scalar fields coexist. The stealthy nature of these black holes may hide important effects during key cosmological epochs, such as inflation or the late-time acceleration of the universe, challenging our ability to observe or model these processes accurately.

In this work, we explore the thermodynamic properties of stealth black hole solutions in various theories, with a focus on scalar-tensor and vector-tensor formulations. Thermodynamics of black holes has emerged as a rich field of study, particularly since it bridges the classical theory of general relativity with quantum aspects of gravity. Key relationships such as the Bekenstein-Hawking entropy, which links a black hole's event horizon area to its entropy, and the laws of black hole thermodynamics, analogous to the laws of classical thermodynamics, suggest deep connections between gravity, quantum mechanics, and thermodynamics. In this context, the study of modified gravity theories and their associated black hole solutions can reveal new insights into how these fundamental concepts might change under different theoretical frameworks. Our analysis primarily investigates how the presence of non-trivial fields influences the thermodynamics of these black holes. A critical aspect of this work is distinguishing between different types of stealth black holes based on whether their additional degrees of freedom are dynamical or algebraic.

Our main findings highlight a fundamental distinction between two types of stealth solutions. First, when the additional degree of freedom, such as a scalar field, does not lead to dynamical equations (in the sense that it does not satisfy a differential equation), it remains non-contributive at the level of boundary terms in the thermodynamic calculations. We will coin this scenario the {\it{bona fide stealth}}. Consequently, the resulting stealth black hole behaves thermodynamically as a classical Schwarzschild or Kerr black hole, without any modifications to the mass or entropy. This implies that such stealth solutions maintain their ``invisibility" not only in their external spacetime geometry but also in their thermodynamic behavior. However, when the scalar or vector fields induce non-trivial boundary terms through their dynamics, the stealth nature is altered at the thermodynamic level. These cases show shifts in the black hole’s thermodynamic quantities, including potential changes in mass and entropy. Specifically, we find that these contributions can lead to modified free energy expressions and adjustments to the first law of thermodynamics for black holes. For instance, in vector-tensor theories, the stealth solution can exhibit an increased free energy compared to the standard Schwarzschild black hole, suggesting a less stable configuration.

The implications of these findings are significant for understanding the nature of stealth solutions. The concept of stealth black holes adds another layer of complexity to the search for deviations from Einstein’s theory of gravity. They suggest that, even in the presence of exotic fields, certain signatures of new physics could remain elusive, avoiding direct detection while still playing a fundamental role in the dynamics of spacetime. This contrasts with scenarios where dynamical contributions from the fields create detectable deviations, potentially allowing for indirect observations of new physics through phenomena like modified gravitational wave signatures during black hole mergers.

This work is organized as follows: In Sec. \ref{thermo}, we analyze the thermodynamic properties of three different stealth black hole scenarios. The two of them belong in the scalar-tensor theories, while the last one is based on a vector-tensor theory. By using Euclidean methods we obtain their mass, entropy and free energy and discuss in detail the possibility of thermodynamic discrepancies between the Schwarschild black hole and the stealth black holes. Finally, Sec. \ref{con} summarizes our results and suggests potential avenues for future research.

\section{Thermodynamic Analysis}\label{thermo}

In this section, we conduct a detailed thermodynamic analysis of various stealth black hole scenarios, examining how these configurations impact fundamental thermodynamic quantities such as mass, entropy, and temperature. Our study begins with the {\it{bona fide stealth}} scenario, where the non-trivial scalar field coexists with the Schwarzschild black hole without modifying its intrinsic thermodynamic properties. We then explore more complex cases where the presence of additional fields leads to shifts in the thermodynamic behavior, revealing how these fields influence boundary terms and contribute to the free energy. This analysis aims to highlight the distinctions between different types of stealth black holes, particularly in how they affect or preserve classical thermodynamic characteristics.

\subsection{Thermodynamics of stealth Schwarzschild (A)dS black hole in scalar-tensor theory}

We begin our analysis by investigating a stealth black hole solution within the framework of Horndeski theory. A key characteristic of this solution is that the functional form of $X \equiv -\partial_{\mu}\phi \partial^{\mu}\phi / 2$ is non-constant. Specifically, we will focus on a class of shift-symmetric Horndeski theories, defined by \cite{Bakopoulos:2023fmv, Baake:2023zsq, Bakopoulos:2023sdm}
\begin{equation}
    S = \frac{1}{2\kappa}\int d^4x \sqrt{-g} \mathcal{L} = \frac{1}{2\kappa}\int d^4x \sqrt{-g} \left\{G_2(X) + G_4(X) R + G_{4X} \left[ (\square \phi)^2 - \phi_{\mu\nu}\phi^{\mu\nu} \right] \right\}~.
\end{equation}
The notation is defined as $\phi_{\mu\nu} = \nabla_{\mu} \nabla_{\nu} \phi$, and subscripts representing differentiation with respect to the corresponding argument. For the coupling functions\begin{equation}
    G_2 = - 2\Lambda + 2\eta \sqrt{X}~,~G_4 = 1+ \lambda \sqrt{X}~,
\end{equation}
a stealth (A)dS-Schwarzschild black hole solution with a dynamical scalar field is obtained
\begin{eqnarray}
    &&ds^2 = -f(r)dt^2 + \frac{dr^2}{f(r)} + r^2 d\Omega^2~,\\
    &&f(r) = 1 - \frac{2M}{r} - \frac{\Lambda r^2}{3}~, \label{sol1}\\
    &&X = \frac{\lambda q^2/2}{\lambda +\eta r^2}~,\\
    &&\phi = q t + \int dr \frac{q}{f(r)}\sqrt{1 - \frac{\lambda f(r)}{\lambda + \eta r^2}}~.\label{sol1a} 
\end{eqnarray}
Note that the linear time dependence of the scalar field is permitted by the shift symmetry of the theory. The constant $q$ is an independent integration constant arising from this symmetry. $M$ represents the geometric mass of the black hole, $\Lambda$ is the cosmological constant, and $\eta$ and $\lambda$ are constants of the theory. Consequently, both $q$ and $M$ are free parameters of the configuration (theory and solution), allowed to vary.

To discuss the thermodynamics of this theory, we utilize the ADM decomposition of a static spacetime with the metric 
\begin{equation}
    ds^2 = N(r)^2 f(r) d\tau^2 + \frac{dr^2}{f(r)} + r^2 d\Omega^2~. \label{adm}
\end{equation}
The partition function for a thermodynamic system will be given by 
\begin{equation}
    \mathcal{Z} = \int d[g_{\mu\nu}^E,\psi]e^{-\mathcal{I}_E}~,
\end{equation}
where the subscript $E$ denotes Euclidean quantities, and $\psi$ collectively represents the matter fields. The spacetime element has an Euclidean signature $(+,+,+,+)$, and we can transition to this signature by performing a Wick rotation $t \to i\tau$. If we are considering a black hole spacetime, then to avoid the conical singularity at the horizon of the black hole $r_h$, the time coordinate $\tau$ must be periodic, with a period given by \cite{Bakopoulos:2024hah,Karakasis:2023hni}
\begin{equation}
    \beta = \frac{4\pi}{\sqrt{g_{\tau\tau}'(r) g_{rr}'(r)}}\Bigg|_{r_h} = \frac{4\pi}{N(r)f'(r)}\Bigg|_{r_h}~, \label{period}
\end{equation}
so that $0 \leq \tau \leq \beta$. This periodicity will be associated with the inverse temperature of the black hole. The key point is that the temperature of the black hole solution will be determined entirely from the spacetime metric. The free energy of a statistical system, $\mathcal{F}$, will be related to the partition function via \cite{Gibbons:1976ue}
\begin{equation}
    \mathcal{F} = -\ln \mathcal{Z}/\beta~. \label{euclid}
\end{equation}
By imposing the saddle point approximation, we may consider that for the black hole spacetime that will concern us in this work, it is sufficient to consider the contribution of the action of the theory evaluated on the classical solution when the field equations hold $\delta \mathcal{I}_E = 0$~. Consequently, we need to calculate the on-shell action of our theory, ensuring it attains a true extremum within the class of fields considered. From this, we can relate the Euclidean action to the free energy of the system.
\begin{equation}
    \mathcal{I}_E = \mathcal{F}\beta~.
\end{equation}
For this calculation, we will use the line element (\ref{adm}) with $\phi = i\chi(t) + \psi(r)$, where $\chi''(t) = 0$ implies $\chi(t) = qt$. Thus, $i\chi(t) = iqt = q\tau = \chi(\tau)$, since only linear time dependence is allowed due to the shift symmetry.

Now, we can write the action using the ADM decomposition as
\begin{equation}
    \mathcal{I}_E = \int d^3x d\tau \left[ P\dot{\phi} -NH \right] + \mathcal{B}~, \label{ac}
\end{equation}
where $P$ is the corresponding conjugate momentum, $N$ is the lapse function, and $H$ is the reduced Hamiltonian. The quantity $\mathcal{B}$ introduces suitable boundary terms to ensure a well-defined variational problem. By utilizing the fact that $\chi''(t) = 0$ and performing integration by parts, we conclude that the reduced Hamiltonian reads
\begin{equation}
    H = \frac{4 \pi  \left(r f'(r)+f(r)+\Lambda  r^2-\sqrt{X(r)} \left(\lambda +\eta  r^2\right)-1\right)}{\kappa }- \frac{2 \pi  \lambda  \chi '(\tau)^2}{\kappa  N(r)^2 \sqrt{X(r)}}~,
\end{equation}
where we have used $X$ and $\chi$ instead of $\chi$ and $\psi$ for simplicity, and we have performed the integrations over the angles. The above equation has been derived by creating total derivatives and canceling boundary terms, utilizing the fact that the theories are equivalent up to boundary terms. Since the quantities in the action (\ref{ac}) are all radial dependent ($\dot{\phi}=q$) we have terms like $\int dr N'(r)\mathcal{W}(r) \sim -\int dr N(r)\mathcal{W}'(r)$, where we have canceled the boundary term $\int dr (d (N(r)\mathcal{W}(r)))/dr$, where $\mathcal{W}$ is a complicated function of $f,X$ and their derivatives. Since only derivatives of the scalar field $\partial_{\mu}\phi$ appear in the action, we are free to substitute $\psi'$ with $X$, as pointed out in \cite{Bakopoulos:2024ogt}.
Using the standard definition for the conjugate momentum, we may identify the momentum of the scalar field as
\begin{equation}
    P(\tau,r)= \frac{1}{\kappa} \frac{\partial \mathcal{L}}{\partial(\partial_ \tau \phi)} = \frac{4 \pi  \lambda  \chi '(\tau)}{\kappa  N(r) \sqrt{X(r)}}~. \label{P}
\end{equation}
Note that $\partial P / \partial \tau = 0$ due to the shift symmetry. Now, we may rewrite the action \ref{ac} in terms of the reduced Hamiltonian and the conjugate momentum as
\begin{equation}
    \mathcal{I}_E = \int_0^{\beta} d \tau \int_{r_h}^{\infty} dr N(r)\left( \frac{\kappa \sqrt{X(r)} P(\tau,r)^2}{8 \pi  \lambda }+\frac{4 \pi \left(r f'(r)+f(r)+\Lambda  r^2-\sqrt{X(r)} \left(\lambda +\eta  r^2\right)-1\right)}{\kappa }\right) + \mathcal{B}~, \label{actionL}
\end{equation}
where we will denote the quantity under the integrals as $\mathcal{H}$. Now, it is time to obtain the field equations, as well as the boundary terms, in order to achieve $\delta \mathcal{I}_E = 0$. The canonical variables of the Euclidean action are $X, f, \chi$, while $N$ is a non-dynamical variable acting as a Lagrange multiplier. Since $X, f, N$ are devoid of momenta, the corresponding field equations read
\begin{eqnarray}
    &&32 \pi ^2 \lambda  \left(\lambda +\eta  r^2\right)-\kappa ^2 P(\tau,r)^2=0~,\\
    &&N'(r)=0~,\\
    &&\frac{\kappa  \sqrt{X(r)} P(\tau,r)^2}{8 \pi  \lambda }+\frac{4 \pi  \left(r f'(r)+f(r)+\Lambda  r^2-\sqrt{X(r)} \left(\lambda +\eta  r^2\right)-1\right)}{\kappa }=0~,
\end{eqnarray}
while Hamilton's equations for $\chi$ read
\begin{eqnarray}
    \frac{\partial \mathcal{H}}{\partial \chi(\tau)}&=&-\partial_\tau P(\tau,r)~,\\
    \frac{\partial \mathcal{H}}{\partial P(\tau,r)}&=&\chi '(\tau)~.
\end{eqnarray}
These two equations are trivially satisfied. The first equation yields $\partial P / \partial \tau = 0$, which is the condition for linear time dependence due to the shift symmetry, while the second returns \ref{P}. Now, returning to the equations for $X, f, N$, we obtain the relations (\ref{sol1}) - (\ref{sol1a}), and one can verify that these configurations lead to the vanishing of the Hamiltonian, $\mathcal{H} \equiv 0$. Consequently, all physical information is encoded in the boundary term $\mathcal{B}$, as now $\mathcal{I} = \mathcal{B}$.

In order to have a well-defined variational principle, we have discarded a boundary term, namely
\begin{equation}
    \delta \mathcal{B} + \beta  \int_{r_h}^{\infty} dr \frac{d}{dr}\left(4\pi r \delta f(r)/\kappa \right) =0~,
\end{equation}
which yields
\begin{equation}
    \delta \mathcal{B} =-  4\pi \beta r \delta f(r) /\kappa \big|_{r_h}^{\infty}~.
\end{equation}
At infinity, the variation of $f$ reads $\delta f = -2 \delta M / r$, while at the horizon, using the period of the Euclidean time coordinate, we have $\delta f(r_h) = -4\pi \delta r_h / \beta$. Utilizing these relations and splitting the variation of the boundary term into two parts, one at infinity and one at the event horizon $\delta \mathcal{B} = \delta \mathcal{B}(\infty) + \delta \mathcal{B}(r_h)$ we obtain 
\begin{equation}
    \delta\mathcal{B}(\infty) = 8\pi \beta \delta M/\kappa ~,
\end{equation}
and 
\begin{equation}
    \delta \mathcal{B}(r_h) = -16\pi^2 r_h \delta r_h/\kappa = -\delta \left(\frac{2\pi A(r_h)}{\kappa}\right) ~,
\end{equation}
and considering the grand canonical ensemble and the black hole situated within a heat bath at a fixed temperature $T \equiv 1/\beta$, we may write
\begin{eqnarray}
    &&\mathcal{B}(\infty) = 8\pi \beta  M/\kappa~,\\
    &&\mathcal{B}(r_h) = - \frac{2\pi A(r_h)}{\kappa}~.
\end{eqnarray}
So, our action reads $\mathcal{I}_E = 8\pi \beta M / \kappa - 2\pi A(r_h) / \kappa$. By comparing this with the free energy of the grand canonical ensemble $\mathcal{F} = \mathcal{M} - T \mathcal{S}$, we may identify $\mathcal{M} = 8\pi M / \kappa$ and $\mathcal{S} = 2\pi A(r_h) / \kappa$ as the mass and entropy of the black hole. The first law of thermodynamics holds by construction, and the free energy of the stealth black hole is identical to that of the Schwarzschild black hole. As a result, these spacetimes cannot be distinguished thermodynamically, thus we have coined them {\it{bona fide stealths}}. On-shell, even at the level of the theory, the stealth scalar field does not affect physics, even though it requires dynamics from its non-minimal coupling to gravity. 

The temperature of the black hole can be solely determined by the spacetime metric; however, the entropy is a byproduct of the theory under consideration, as well as the conserved mass of the black hole. These quantities depend on the effective Newton's constant. The shift symmetry of the theory renders the entropy independent of the value of the scalar field at the horizon of the black hole, since this value is arbitrary (we can set $\phi(r_h) \to \phi(r_h) + c$, where $c$ is a constant, without any physical consequence). However, this would change the value of the entropy by a constant, and since entropy is a physical quantity measuring the disorder of the system (theory and solution), we would have different values for the entropy for the same system, which is non-physical. Nevertheless, the entropy does not acquire even a correction term depending on the horizon of the black hole, while terms ``allowed" by the shift symmetry, such as the kinetic term $X=-\partial_{\mu}\phi\partial^\mu\phi/2$ do not affect the entropy either. 

As we have already discussed, the free energy is the same as in the Schwarzschild case, which means that both the stealth Schwarzschild black hole arising from Horndeski theory and the Schwarzschild black hole for $\phi = 0$ will have the same on-shell actions and are therefore equally probable configurations at the thermodynamic level. This effectively means that the scalar field does not contribute on-shell at all. We have verified this by calculating the action for $G_2 = 2\eta \sqrt{X}$ and $G_4 = \lambda \sqrt{X}$, and we found it to be vanishing for the form of the solution given in equations (\ref{sol1} - \ref{sol1a}). As a result, the kinetic energy of the scalar field acquires particular dynamics such that the solution will be the Schwarzschild-(A)dS black hole at the level of the action. This behavior, where the on-shell action of the additional fields vanishes for stealth black holes, with the field acquiring dynamics in a way that does not contribute on-shell, is also noted in the context of wormholes \cite{Bakopoulos:2023tso}.

\subsection{Galileon stealth} \label{galileon}

We will now move on to an example where we obtain modified thermodynamic quantities. 
This example involves the first stealth black hole solutions within the framework of Horndeski theory, as explored in the seminal works of \cite{Babichev:2013cya, Kobayashi:2014eva}.  The action functional of the theory has the form
\begin{equation}
    S = \int d^4x \sqrt{-g} \left( \zeta R +\alpha G_{\mu\nu}\nabla^{\mu}\phi\nabla^{\nu}\phi\right)~, \label{stealthbabi}
\end{equation}
which is essentially a Horndeski theory with $G_4(X) = \zeta + \beta X$, and as a result, the action can be expressed as \cite{Kobayashi:2011nu}
\begin{equation}
    S = \int d^4x \sqrt{-g} \left( (\zeta + \alpha X) R +\alpha ((\square{\phi})^2 - \nabla_{\mu}\nabla_{\nu}\phi\nabla^{\mu}\nabla^{\nu}\phi)\right)~. \label{stealthbabi1}
\end{equation}
The above theories have as a solution the Schwarzschild black hole, with,
\begin{equation}
    f(r) = 1-\frac{2M}{r}~, \label{schwa}
\end{equation}
in (\ref{sol1}), with a non-trivial, linearly time-dependent scalar field: $\phi = qt + \psi(r)$, where $q$ is an independent integration constant related to the shift symmetry of the theory. However, in this case, the resulting kinetic term is constant, $X \equiv - (\partial \phi)^2 / 2 = q^2 / 2$, unlike the previous case. 
Let us discuss explicitly the fact that the energy momentum tensor arising when varying with respect to the metric indeed vanishes, using the theory (\ref{stealthbabi}). Utilizing the metric (\ref{adm}) we can obtain the components of the energy-momentum tensor as:
\begin{eqnarray}
    && T^{\tau\tau} = \frac{\alpha  \left(N q^2 \left(r f'-1\right)+f N^3 \left(X \left(r f'+f+1\right)+2 f r X'\right)+2 f q^2 r N'\right)}{f^2 N^5 r^2}~,\\
    && T^{\tau r} = -\frac{i \alpha  q \sqrt{q^2-2 f N^2 X} \left(N \left(r f'+f-1\right)+2 f r N'\right)}{f N^4 r^2}~,\\
    && T^{rr} = \frac{\alpha  \left(-N q^2 \left(r f'+2 f-1\right)+f N^3 X \left(3 r f'+3 f-1\right)+6 f^2 N^2 r X N'-2 f q^2 r N'\right)}{N^3 r^2}~,\\
    && T^{\theta\theta} = \frac{\alpha  \left(N^2 \left(N' \left(3 r X f'+2 f \left(r X'+X\right)\right)+N \left(X' \left(r f'+2 f\right)+X \left(r f''+2
   f'\right)\right)+2 f r X N''\right)+2 q^2 N'\right)}{2 N^3 r^3}~.
\end{eqnarray}
The non-diagonal component arises due to the linear-time-dependent scalar field. Substituting the metric (\ref{schwa}), alongside $N=1$ we can see that the non-diagonal component vanishes, while the other equations will yield a relation for $X$, when we set them equal to $0$. Remember that the Einstein tensor vanishes by definition for the Schwarzschild black hole. From  $T^{\theta\theta}=0$ we obtain $X'=0$, while from $T^{rr}=0$ we obtain 
\begin{equation}
    X = q^2/2~,
\end{equation}
which satisfies the $T^{\tau\tau}=0$ condition. Hence this is a stealth solution, as defined by the system ({\ref{stealthcondition}), when $\Lambda=0$.} 
The specific form of $\psi(r)$ is irrelevant to the following arguments, so we refer to \cite{Babichev:2013cya, Kobayashi:2014eva} for further details. 

Here, we will not present the detailed calculations, as the procedure is similar to the previous case, where we nullify the reduced Hamiltonian, and consequently, the Euclidean action is determined by the boundary term. We are now in a position to calculate the variation of the boundary term to establish a well-defined variational procedure, $\delta \mathcal{I}_E = 0$. The free parameters allowed in the variation are the mass $M$ and the scalar charge $q$. In \cite{Minamitsuji:2023nvh}, the authors stated that the parameter "$q$ is not the integration constant but the constant appearing in the ansatz of the scalar field compatible with the shift symmetry..."; however, by introducing an auxiliary function $\chi(t)$ for the time dependence of the scalar field, as we did in the previous case and as shown in \cite{Bakopoulos:2023sdm, Bakopoulos:2024ogt}, one can derive a first-order differential equation for $\chi(t)$ from the requirement of spacetime staticity, resulting in a linearly dependent scalar field. Thus, $q$ is indeed an integration constant allowed in the variations. Of course, one must impose the constraint $\chi''(t) = 0$, which is necessary to ensure a finite action and a static black hole spacetime \cite{Babichev:2013cya}. Starting from the theory (\ref{stealthbabi}), to compute the field equations, we ignored the following boundary contributions:
\begin{equation}
    -\frac{16 \pi  \alpha \delta X f r}{T}-\frac{8 \pi 
   \delta f r (\alpha  X-\zeta )}{T}\bigg|_{r_h}^{\infty}+\delta\mathcal{B}=0~.
\end{equation}
Notice that the first term contains the variation of the kinetic energy of the scalar field, while the second term includes an additional contribution arising from the scalar field. These contributions would be absent if we were dealing with the Schwarzschild black hole in the context of general relativity. Splitting the variation of the boundary term into two parts, one at infinity and one at the event horizon $\delta \mathcal{B} = \delta \mathcal{B}(\infty) + \delta \mathcal{B}(r_h)$, we have the following expression at infinity:\begin{equation}
    \delta\mathcal{B}(\infty) + \frac{8 \pi  (\alpha  q (4 \delta q M+ \delta M q-2 \delta q r)-2 \delta M \zeta )}{T}=0~,
\end{equation}

where $T$ is the temperature of the black hole. To ensure a finite boundary term at infinity, we must set $\delta q = 0$, thereby canceling the divergent terms. This restriction on the variation of $q$ is not because $q$ is not an integration constant, but rather due to the requirement of finiteness in the boundary contributions. Consequently, we obtain: 
\begin{equation}
    \delta\mathcal{B}(\infty) =\frac{8 \pi  \delta M \left(2 \zeta - \alpha  q^2 \right)}{T}~.
\end{equation}
Now, at the event horizon, keeping in mind that $\delta q=0$ we have 
\begin{equation}
    \delta \mathcal{B}(r_h) = 16 \pi ^2 \delta r_h r_h \left(\alpha  q^2-2 \zeta \right)  ~
\end{equation}
Dropping the variations and considering the Grand Canonical Ensemble, we obtain
\begin{eqnarray}
    &&\mathcal{B}(\infty) = \frac{8 \pi  M \left(2 \zeta-\alpha  q^2 \right)}{T}~,\\
    &&\mathcal{B}(r_h) = -2 \pi A(r_h) \left(2 \zeta - \alpha  q^2\right)~,
\end{eqnarray}
where $A(r_h) = 4\pi r_h^2$. The Euclidean action and therefore the boundary term will be related to the free energy of the system via (\ref{euclid}), so we can obtain the conserved mass and entropy of the black hole as
\begin{equation}
    \mathcal{M} = 8 \pi  M \left(2 \zeta-\alpha  q^2 \right)~, \hspace{0.5cm} \text{and} \hspace{0.5cm} \mathcal{S} = 2 \pi A(r_h) \left(2 \zeta - \alpha  q^2\right)~. \label{entropygalileon}
\end{equation}

Now we will discuss the thermodynamics of this solution using Euclidean arguments based on the action (\ref{stealthbabi1}). Following the same procedure, we eventually obtain the same boundary terms and, as a result, the same conserved mass and entropy for the black hole solution. Therefore, at the on-shell level, since the theories (\ref{stealthbabi}) and (\ref{stealthbabi1}) are equivalent up to total divergences, they yield the same field equations, and consequently, the same boundary terms arise from their variation. Notice also the shifted entropy formula in this particular example. In the Appendix, we have included the calculation of the entropy using the well-known Wald formula for entropy \cite{Wald:1993nt}. We performed the calculation for the theory (\ref{stealthbabi}) and found the classic Area Law for entropy. Similar ambiguities have been noted previously \cite{Jacobson:1993vj,Hajian:2020dcq} (for more details, we refer to the Appendix). Nonetheless, the Euclidean calculation of thermodynamic quantities yields consistent results, based on the fundamental relation of statistical physics and quantum field theory for finite-temperature systems (\ref{euclid}), along with the need for a well-defined variational procedure, which is ensured by choosing appropriate boundary contributions. Therefore, we rely on 
the Euclidean calculations.

In this case, the free energy of the stealth Schwarzschild solution will be given by
\begin{equation}
    \mathcal{F}_{ST} = \frac{1}{4} r_h \left(1-8 \pi  \alpha  q^2\right)
\end{equation}
where, for simplicity, we have set  $\zeta=1/16\pi$ effectively setting Newton's constant to unity, while the free energy of the Schwarzschild black hole will be given by $\mathcal{F}_{GR} = r_h/4$. Now, we can compare these two solutions. To do so, we have to position these black holes in the same heat bath. Since the thermodynamics of heat bath fixes the temperature, and as a result the event horizon of the black hole, in general, the black holes to be compared have to have the same horizon radius. However, in our case the black holes are already in the same heat bath, having the same temperature. The free-energy difference between the two configurations now reads
\begin{equation}
    \Delta \mathcal{F} = \mathcal{F}_{GR}- \mathcal{F}_{ST} = -2 \pi  \alpha  q^2 r_h~,
\end{equation}
and is therefore dependent on the sign of the coupling parameter $\alpha$. A negative $\alpha$ will imply that the stealth solution requires less free energy and thus is a more probable configuration, implying thermodynamic prefer-ability, while a positive $\alpha$ hints that the classical Schwarzschild black hole is the one with less free energy and thus thermodynamically preferred. Regarding the local thermodynamic stability, by calculating the heat capacity $C(r_h) = TdS/dT = 2 \pi  r_h^2 \left(8 \pi  \alpha  q^2-1\right)$ one can see that for a negative $\alpha$ the stealth black hole still behaves like the Schwarzschild black hole, while for $\alpha>(8\pi q^2)^{-1}$ the heat capacity will be positive, implying thermodynamic stability, however, this case is ill-defined since we encounter both negative mass and negative entropy and as a result, gravity becomes repulsive, 
however, this case is ill-defined since we encounter both negative mass and negative entropy.
Note that the true conserved mass is given in equation (\ref{entropygalileon}) and $M$ acts as an integration constant, not a conserved quantity, therefore a negative conserved mass will imply repulsive gravity.

It is clear that a negative mass will imply a repulsive gravity, but it can also be seen as follows. The entropy of a black hole is, in general, given by \cite{Brustein:2007jj}

\begin{equation}
   \mathcal{S} = \frac{A   (r_h)}{4G_{\text{eff}}}~,  
\end{equation}

where here $G_{\text{eff}}$ will be a scale dependent quantity since it will depend on the constants of the theory and, more importantly, on the event horizon radius. Now, since $A(r_h)>0$ the sign of the entropy is determined by the sign of $G_{\text{eff}}$ and a negative $G_{\text{eff}}$ will imply repulsive gravity. Here, in this particular example, $G_{\text{eff}}$ does not depend on the event horizon radius, it instead depends on the constants of the theory and $q$ and hence these quantities will determine the nature of gravity. 
As one can see, the stealth field introduces major challenges and intricacies at the thermodynamic level, that are undetectable at the level of the metric element.
Here lies the major point of our work. In the cases where the additional degree of freedom is not dynamical, in the sense that it does not obey a differential equation but an algebraic one, as the {\it{bona fide stealth}}, we do not obtain boundary terms and as a result, only GR contributes. So this kind of stealth black holes, preserve the stealth nature at the thermodynamic level. In any case where boundary terms emanate from the additional stealth fields, contributions to the thermodynamic quantities are expected, even shifts in the mass and entropy as in this case.

\subsection{Thermodynamics of stealth Schwarzschild black hole in vector-tensor theory} \label{GAA}
In our final example, we will investigate a different case of a stealth black hole solution, this time derived from a vector-tensor theory. A Schwarzschild black hole solution can be identified as a stealth solution within the framework of the following theory:

\begin{equation}
    S = \int d^4x \sqrt{-g} \left( \frac{1}{16\pi} R - \frac{1}{4} F_{\mu\nu}F^{\mu\nu} + \gamma G^{\mu\nu}A_{\mu}A_{\nu}\right)~, \label{stealthvector}
\end{equation}
where $F_{\mu\nu} = \partial_{\mu}A_{\nu} - \partial_{\nu}A_{\mu}$ is the field strength and $A_{\mu}$ is the vector field. For $\gamma=1/4$, one can obtain the following stealth configuration \cite{Chagoya:2016aar,Minamitsuji:2016ydr}
\begin{eqnarray}
    &&f(r) = 1- \frac{2M}{r}~, \label{sol2a}\\
    &&A_{\mu} = (A_0(r),A_1(r),0,0)~,\\
    &&A_0(r) = Q/r +P~,\\
    &&A_1(r) = \frac{\sqrt{2 P r (M P+Q)+Q^2}}{r-2 M}~. \label{sol2d}
\end{eqnarray}
To perform the thermodynamic analysis we will rewrite the action (\ref{stealthvector}) in Hamiltonian form, using the ADM decomposition in the Euclidean sector (\ref{adm}). Here, we will show the relevant calculations, in order to discuss some important points that arise.

The total action then will read
\begin{eqnarray}
    &&\mathcal{I}_E = \mathcal{H}_{EM}+ \mathcal{H}_{GRAV}+ \mathcal{B}_E~, \label{euclideanaction1}\\
    &&\mathcal{H}_{EM} = \int d^3x d\tau\left(p^{\mu}\partial_{0}A_{\mu} - \mathcal{L}_{EM}\right)~,\\
    &&\mathcal{H}_{GRAV} = -\int d^3xd\tau NH~,
\end{eqnarray}
where $\mathcal{B}_E$ will take care of both the electromagnetic as well as the gravitational boundary terms. Performing the integration over the periodic coordinates and after integration by parts, the gravitational contribution reads
\begin{equation}
    \mathcal{H}_{GRAV} = \frac{1}{2} \beta\int dr   N(r) \left(r f'(r)+f(r)-1\right)~. \label{hgrav}
\end{equation}
The conjugate momenta for the electromagnetic field of our theory yields 
\begin{equation}
    p^\mu = -\sqrt{|g|}F^{0\mu}~,
\end{equation}
and is unaffected by the coupling of the Einstein tensor with the gauge field because of its lacking in derivative coupling. For our ADM metric this yields
\begin{equation}
    p^r = \frac{r^2}{N(r)}A'_0(r)~.
\end{equation}
The Hamiltonian for the electromagnetic part becomes 
\begin{equation}
    \mathcal{H}_{EM} = 4\pi\beta\int dr \left(-\partial_{\mu}p^{\mu}A_0 +  N \left( A_1 \gamma  f \left(A_1 \left(3 r f'+f+1\right)+4 A_1' f r\right)-\frac{(p^{r})^{2}}{2r^2}\right)-\frac{
     A_0^2  \gamma  \left(r f'+f-1\right)}{f N} \right) \label{hem}
\end{equation}
The Hamiltonian version has the advantage that, since our solution is static, it can be written as a linear combination of constraints that vanish when the field equations hold. Indeed, one can see that variation with respect to $N$ in (\ref{hgrav}) will yield the Schwarszchild metric and so $H_{GRAV}$ is null on-shell. As a result, the variation of the total Hamiltonian with respect to the dynamical fields $N,f,A_1,A_0,p^{r}$ results in the following field equations
\begin{eqnarray}
    &&\frac{1}{2} \beta  \left(\frac{2 \pi  A_0^2 \left(r f'+f-1\right)}{f N^2}+2 \pi  A_1 f^2 \left(4 r A_1'+A_1\right)+2
   \pi  A_1^2 f \left(3 r f'+1\right)+r f'+f-\frac{4 \pi  p^2}{r^2}-1\right)=0~,\\
   &&\frac{1}{2} \beta  \left(\frac{2 \pi  A_0^2 \left((f-1) N-f r N'\right)}{f^2 N^2}+\frac{4 \pi  A_0 r A_0'}{f N}-2 \pi 
   A_1^2 \left((f-1) N+3 f r N'\right)+4 \pi  A_1 f N r A_1'-r N'\right)=0~,\\
   &&-2 \pi  A_1 \beta  f \left(N \left(r f'+f-1\right)+2 f r N'\right)=0~,\\
   &&\frac{A_0 \beta  \left(r f'+f-1\right)}{f N}+2 \beta  (p^r)'=0~,\\
   &&4 \pi  \beta  \left(A_0'-\frac{N p^r}{r^2}\right)=0~.
\end{eqnarray}
These equations admit the solution given in equations (\ref{sol2a})-(\ref{sol2d}) with $N=1$, provided that we Wick-rotate the integration constants $Q,P$ in the Euclidean sector: $Q\to i Q,~ P\to i P$, thus generating solutions with complex constants in real equations, as done for the BTZ black hole case \cite{Banados:1992wn}. Then the momentum for the electromagnetic field on-shell in the Euclidean sector becomes imaginary $p^r = -i Q$. For these configurations, the total Hamiltonian in (\ref{euclideanaction1}) vanishes on shell and all physical information is encoded in the boundary term, $\mathcal{B}_E$. In order to obtain the field equations, we canceled the following boundary terms
\begin{equation}
   \left(\delta \mathcal{B}_E + \frac{1}{2} \beta  \left(\frac{2 \pi  A_0^2 \delta f r}{f N}-8 \pi  A_0 \delta p+N r (2 \pi  A_1 f (4
   \delta A_1 f-3 A_1 \delta f)+\delta f)\right)\right)\Bigg|_{r=r_h}^{\infty}=0~. \label{boundary}
\end{equation}
Taking into account the solution of the field equations, we may calculate the variation of the fields at the boundaries, keeping in mind that, at least at the moment, $M,P,Q$ are free to vary. Then, at infinity we find
\begin{equation}
    \delta \mathcal{B}_E(\infty) =  -\beta  \left(\delta M \left(2 \pi  P^2-1\right)+4 \pi  \delta P (2 M P+Q)\right)~.
\end{equation}
At the horizon we have that $f(r_h) \sim 0 + f'(r_h)(r-r_h) \to \delta f = -f'(r_h)\delta r_h$. As can be observed from (\ref{boundary}), there is a potentially problematic divergent term present, specifically:
\begin{equation}
\lim_{r\to r_h}\left(\frac{2 \pi  A_0^2 \delta f r}{f N}\right) \to \infty~,
\end{equation}
will diverge since $f(r_h)=0$ by definition of the event horizon.
To ensure finite contributions at the boundary, it is necessary to restrict the free parameters of the system. In particular, we must set $A_0(r_h)=0$, which implies $P=-Q/r_h = -Q/(2M)$. Consequently, the variation of $P$ is governed by the variation of the constants $M$ and $Q$ via
\begin{equation}
    \delta P = \frac{\delta M Q}{2 M^2}-\frac{\delta Q}{2 M}~. \label{deltaP}
\end{equation}
In light of the above considerations, the boundary term at the event horizon is expressed as:%
\begin{equation}
    \delta \mathcal{B}_E(r_h) = -2\pi r_h \delta r_h = - \pi \delta r_h^{2} = -\frac{\delta A(r_h)}{4}~,
\end{equation}
where $A(r_h) = 4\pi r_h^2$ is the area of the black hole, while the restriction on $P$ results in 
\begin{equation}
    \delta \mathcal{B}_E(\infty) = \beta  \delta M - \frac{\pi  \beta   Q^2}{2 M^2} \delta M~, \label{infbound}
\end{equation}
for the boundary term at infinity. $Q$ is allowed in the variation, as is clear from equation (\ref{deltaP}), however, due to the condition of finiteness at the event horizon, there is no contribution from $\delta Q$ at the boundary.  Considering the Grand Canonical ensemble while keeping the temperature fixed, we can drop the variations, yielding:
\begin{eqnarray}
    &&\mathcal{B}_E(\infty) = \beta  M+\frac{ \pi  \beta  Q^2}{2 M}~,\\
    &&\mathcal{B}_E(r_h) = -A(r_h)/4~.
\end{eqnarray}
Comparing with the free energy we may identify the total conserved mass of the spacetime as $\mathcal{M} = M+\frac{ \pi    Q^2}{2 M}$ and the entropy of the black hole as $\mathcal{S} = A(r_h)/4$.
It is therefore evident that despite having the same horizon radius $r_h=2M$ the stealth black hole possesses higher free energy when compared to the classic Schwarzschild black hole and the Euclidean action (free energy $\mathcal{F}$) of this theory will be given by
\begin{equation}
    \mathcal{I}_E = \beta \mathcal{F} = \beta\left( M+\frac{ \pi    Q^2}{2 M} -\frac{1}{\beta} \frac{A(r_h)}{4}\right)~.
\end{equation}
Moreover, this difference will be significant for small black holes (possibly primordial black holes), whereas for astrophysical black holes formed from stellar collapse, the difference will be negligible. 
As a result, the stealth Schwarzchild black hole in this theory admits a modified first law of thermodynamics, namely 
\begin{equation}
    \delta \mathcal{M} = T \delta \mathcal{S}~, \label{firstlaw}
\end{equation}
in order to have a well-defined variational procedure. Concluding, the expression for the total conserved mass has been obtained by considering that $M$ is allowed to vary, as in the Schwarzschild scenario. Although $Q$ did not appear directly in the calculations, its relationship with $P$ played a pivotal role in obtaining finite contributions at the horizon boundary.

 The difference in free energies between the Schwarzschild and the stealth black hole is then
\begin{equation}
    \Delta \mathcal{F} = \mathcal{F}_{GR} - \mathcal{F}_{ST} = (M-T\mathcal{S}) - (\mathcal{M}-T\mathcal{S}) = M-\mathcal{M} = - \frac{ \pi   Q^2}{2 M}<0~,
\end{equation}
which implies that the GR Schwarzschild solution has lower free energy and is therefore a more stable and probable configuration than the stealth configuration. In other words, considering the stealth action and the stealth black hole, there exists a non-zero probability that the stealth black hole will decay to the Schwarzschild black hole spacetime.
To assess local thermodynamic stability, we need to use the specific heat, which is defined as:
\begin{equation}
    C \equiv T\frac{d\mathcal{S}}{dT} \equiv \frac{d\mathcal{M}}{dT}~,
\end{equation}
where in the second equality we have used the first law of thermodynamics (\ref{firstlaw}). Since $M$ is the geometric mass, we can replace it with the event horizon radius yielding $M=r_h/2$ while the heat capacity becomes
\begin{equation}
    C = -2\pi r_h^2 + 4\pi^2 Q^2~.
\end{equation}
where of course for $Q=0$ it reduces to the Schwarzschild black hole. As evident from the above relation, the heat capacity may turn out to be positive when $Q^2>4M^2/(2\pi)$, which is of course not the case for the classic Schwarzschild solution where the heat capacity is always negative, implying the local thermodynamic instability of the black hole. Moreover, there exists a zero point in the heat capacity, namely $r_{\text{crit}} = \sqrt{2 \pi } Q$. This critical value corresponds to an extremum, particularly a minimum, in the conserved mass. Since we have a vanishing heat capacity, this implies that the black hole cannot absorb or release energy; in other words, adding or removing heat energy from the black hole will not affect its temperature. Zero heat capacity can be encountered at the final stage of evaporation when the black hole becomes extremal \cite{Bakopoulos:2024ogt}. However, in this case, we observe a finite, non-zero temperature, suggesting that we encounter a phase transition, which is absent in the classic Schwarzschild black hole. As a result for $r_h>r_{\text{crit}}$ the stealth black hole behaves similarly to the Schwarzschild black hole, while sufficiently small stealth black holes can be in thermal equilibrium with a heat bath. This observation aligns with the understanding that, in general, small black holes are significantly affected by this shift in the conserved mass compared to larger black holes.

\section{Conclusions}\label{con}

Stealth black hole solutions have emerged as a crucial concept in the study of modified gravity theories, offering a unique challenge to our understanding of how deviations from general relativity might manifest. These solutions allow non-trivial fields, such as scalar or vector fields, to coexist with classical black hole geometries like Schwarzschild or Kerr, without altering the observable external spacetime. This ``stealthy" behavior means that, despite the presence of additional degrees of freedom, the black hole appears identical to its general relativistic counterpart in terms of mass, spin, and other observable properties. This has significant implications for the ongoing efforts to test general relativity in strong gravitational regimes, as it suggests that some modifications to gravity might remain hidden from standard observational techniques.

The importance of understanding stealth solutions lies in their potential to reveal limitations in our current methods of probing gravity. Traditional tests of general relativity often rely on detecting discrepancies between observed gravitational phenomena and theoretical predictions. However, the existence of stealth solutions means that modified gravity theories can evade these tests, potentially leading to scenarios where deviations from general relativity exist but remain undetected. This complicates the search for a unified theory of gravity, as it suggests that new physics could be present without leaving a clear imprint on black hole dynamics or astrophysical observations.

In this work, we investigated the thermodynamic properties of stealth black holes in various modified gravity theories, with a focus on scalar-tensor and vector-tensor formulations. Our analysis centered on understanding how the presence of non-trivial fields, such as scalar or vector fields, influences the thermodynamic behavior of black holes. We identified key distinctions between types of stealth solutions based on whether the additional degrees of freedom are dynamical or algebraic. This differentiation is crucial for assessing the extent to which modifications to general relativity can remain hidden at the level of black hole observations.

One of the central results of our study is the identification of a class of stealth solutions where the additional fields do not alter the thermodynamics of the black hole. These ``genuine" stealth black holes retain their stealthy nature at the thermodynamic level, meaning that their mass, entropy, and other thermodynamic quantities are identical to those of their general relativistic counterparts. This occurs because the non-trivial fields in these solutions do not lead to boundary terms that contribute to the on-shell action, resulting in a complete thermodynamic match with classical black hole solutions. As a result, these configurations are entirely indistinguishable from Schwarzschild or Kerr black holes, both geometrically and thermodynamically.

In contrast, when the additional degrees of freedom become dynamical, they contribute to boundary terms that alter the black hole’s thermodynamic quantities. In such cases, we found that modifications to the free energy, shifts in mass, and changes to the entropy can arise, reflecting the presence of the non-trivial fields. This distinction allows us to classify stealth black holes into different categories based on their thermodynamic signatures, providing a framework for understanding how new degrees of freedom in modified gravity theories might manifest.

In conclusion, our study sheds light on the subtle and intricate ways that modified gravity theories can manifest through black hole thermodynamics.

\section{Acknowledgements}
\noindent The research project was supported by the Hellenic Foundation for Research and Innovation (H.F.R.I.) under the “3rd Call for H.F.R.I. Research Projects to support Post-Doctoral Researchers” (Project Number: 7212). A.B. also acknowledges participation in the COST Association Action CA21136 “Addressing observational tensions in cosmology with systematics and fundamental physics (CosmoVerse)”. We acknowledge useful discussions with Panos Dorlis and Cristi\'an Erices.

\appendix
\section{Confirmation of the Area Law using Wald's entropy formula}

Here, we will present the calculation of the entropy for the theory (\ref{stealthbabi}) via the Wald formula. 
The Wald entropy is given by \cite{Wald:1993nt,Iyer:1994ys}
\begin{equation}
\mathcal{S} = -2\pi \oint d^2x \sqrt{h} \frac{\partial \mathcal{L}}{\partial R_{\alpha\beta\gamma\delta}} \hat{\epsilon}_{\alpha\beta}\hat{\epsilon}_{\gamma\delta}~,
\end{equation}
where the integral is evaluated at the event horizon of the black hole, and $\hat{\epsilon}_{\alpha\beta} = \xi_{\alpha}\eta_{\beta} - \xi_{\beta}\eta_{\alpha}$ is the binormal vector to the horizon surface. $\xi^{\alpha}$ is the null Killing vector generating the horizon and $\eta^{\alpha}$ is the null normal vector on the horizon \cite{PhysRevD.74.044007,Tachikawa:2006sz}. Therefore we take $\xi_\alpha = -f(r)dt$ and $\eta_\alpha = -dt + dr/f(r)$.

The Einstein-Hilbert (E-H) term in general gives 
\begin{equation}
\frac{\partial R}{\partial R_{\alpha\beta\gamma\delta}}\hat{\epsilon}_{\alpha\beta}\hat{\epsilon}_{\gamma\delta}=
\frac{\partial (g^{\mu\nu}g^{\kappa\sigma}R_{\sigma\mu\kappa\nu})}{\partial R_{\alpha\beta\gamma\delta}}\hat{\epsilon}_{\alpha\beta}\hat{\epsilon}_{\gamma\delta} = g^{\mu\nu}g^{\kappa\sigma}\delta_{\sigma}^{\alpha}\delta_{\mu}^{\beta}\delta_{\kappa}^{\gamma}\delta_{\nu}^{\delta}\hat{\epsilon}_{\alpha\beta}\hat{\epsilon}_{\gamma\delta} = -2~.
\end{equation}
due to the normalization of the binormals $\hat{\epsilon}_{\alpha\beta}\hat{\epsilon}^{\alpha\beta}=-2$. 
Taking into consideration now the integration at the null surface of the horizon, one can obtain that $\mathcal{S}_{\text{E-H}} \sim A(r_h)$, with $A(r_h)=4\pi r_h^2$.
In our case the entropy of the black hole configuration will be schematically given by 
\begin{equation}
    \mathcal{S} = \mathcal{S}_{\text{E-H}} + \mathcal{S}_{\phi} = \mathcal{C}A(r_h) + \mathcal{S}_{\phi}~,
\end{equation}
where $\mathcal{C}$ is related to the Newton constant of the theory at hand. The goal here is to compute the $\mathcal{S}_{\phi}$.

\subsection{Wald entropy for the Galileon stealth}
In the case of a curvature coupling such as the Einstein-kinetic coupling, the entropy may receive corrections. Here we have to deal with the quantity: $\beta G_{\mu\nu}\nabla^{\mu}\phi\nabla^{\nu}\phi=\beta(R_{\mu\nu}\nabla^{\mu}\phi\nabla^{\nu}\phi - R\nabla^{\mu}\phi\nabla_{\mu}\phi/2)$. For notational convenience, we will denote the entropy correction coming from the Ricci tensor term as $\mathcal{S}_{R_{\mu\nu}}$ and the entropy correction coming from the Ricci scalar term as $\mathcal{S}_R$. For the Ricci tensor term one obtains by taking the derivative with respect to the Riemann tensor $g^{\alpha\gamma}A^\beta A^\delta \hat{\epsilon}_{\alpha\beta}\hat{\epsilon}_{\gamma\delta}$ and now for the aforementioned definition of the binormal vector, the Ricci tensor term gives 
\begin{equation}
\mathcal{S}_{R_{\mu\nu}}=8\pi^2 \beta r_h^2 g^{\alpha\gamma}\nabla^{\beta}\phi\nabla^{\delta}\phi~\hat{\epsilon}_{\alpha\beta}\hat{\epsilon}_{\gamma\delta} = -8 \pi ^2 \beta   q^2 r_h^2~.
\end{equation}
Now for the Ricci scalar term we have 
\begin{equation}
\mathcal{S}_{R}=-16\pi^2 \beta r_h^2 \left((-X)/2\right)= 8 \pi ^2 \beta   q^2 r_h^2~,
\end{equation}
and hence 
\begin{equation}
    \mathcal{S_{\phi}} = \mathcal{S}_{R_{\mu\nu}} + \mathcal{S}_{R}=0~.
\end{equation}

As one can see there are no correction terms correcting the Area Law of the entropy that survive at the end, and therefore the entropy will be given by the Bekenstein-Hawking Area Law \cite{Bekenstein:1973ur,Hawking:1975vcx}, while our method clearly establishes a shift in the entropy. This behavior has been pointed out also in \cite{Feng:2015oea}, where the full Wald procedure and the Wald formula yield different results. This feature has been attributed to the fact that $\nabla_{\mu}\phi$ diverges at the event horizon of the black hole. As pointed out in \cite{Jacobson:1993vj}, despite the fact that, scalars and their derivatives formed by the matter fields in the action will not diverge at the bifurcation surface of the event horizon where the entropy is evaluated, a tensor field might diverge, even if all scalars obtained from it are regular there.  Hence, the Wald formula may not be the best way to obtain the entropy of these solutions. For this reason, we relied on Euclidean calculations, where the emergence of thermodynamic quantities is clear and robust.

\bibliography{Refs}

\providecommand{\href}[2]{#2}\begingroup\raggedright\begin{thebibliography}{10}

\bibitem{Ayon-Beato:2004nzi}
E.~Ayon-Beato, C.~Martinez, and J.~Zanelli, ``{Stealth scalar field overflying
  a (2+1) black hole},''
  \href{http://dx.doi.org/10.1007/s10714-005-0213-x}{{\em Gen. Rel. Grav.}
  {\bfseries 38} (2006) 145--152},
  \href{http://arxiv.org/abs/hep-th/0403228}{{\ttfamily arXiv:hep-th/0403228}}.

\bibitem{Banados:1992wn}
M.~Banados, C.~Teitelboim, and J.~Zanelli, ``{The Black hole in
  three-dimensional space-time},''
  \href{http://dx.doi.org/10.1103/PhysRevLett.69.1849}{{\em Phys. Rev. Lett.}
  {\bfseries 69} (1992) 1849--1851},
  \href{http://arxiv.org/abs/hep-th/9204099}{{\ttfamily arXiv:hep-th/9204099}}.

\bibitem{Hassaine:2013cma}
M.~Hassaine, ``{Rotating AdS black hole stealth solution in D=3 dimensions},''
  \href{http://dx.doi.org/10.1103/PhysRevD.89.044009}{{\em Phys. Rev. D}
  {\bfseries 89} no.~4, (2014) 044009},
  \href{http://arxiv.org/abs/1311.4623}{{\ttfamily arXiv:1311.4623 [hep-th]}}.

\bibitem{Alvarez:2016qky}
A.~Alvarez, C.~Campuzano, M.~Cruz, E.~Rojas, and J.~Saavedra, ``{Stealths on
  $(1+1)$-dimensional dilatonic gravity},''
  \href{http://dx.doi.org/10.1007/s10714-016-2158-7}{{\em Gen. Rel. Grav.}
  {\bfseries 48} no.~12, (2016) 165},
  \href{http://arxiv.org/abs/1611.03022}{{\ttfamily arXiv:1611.03022 [gr-qc]}}.

\bibitem{Ayon-Beato:2005yoq}
E.~Ayon-Beato, C.~Martinez, R.~Troncoso, and J.~Zanelli, ``{Gravitational
  Cheshire effect: Nonminimally coupled scalar fields may not curve
  spacetime},'' \href{http://dx.doi.org/10.1103/PhysRevD.71.104037}{{\em Phys.
  Rev. D} {\bfseries 71} (2005) 104037},
  \href{http://arxiv.org/abs/hep-th/0505086}{{\ttfamily arXiv:hep-th/0505086}}.

\bibitem{Ayon-Beato:2024xgp}
E.~Ay\'on-Beato, M.~Hassaine, and P.~A. S\'anchez, ``{Non-Noetherian Conformal
  Cheshire Effect},'' \href{http://arxiv.org/abs/2408.00086}{{\ttfamily
  arXiv:2408.00086 [hep-th]}}.

\bibitem{Caldarelli:2013gqa}
M.~M. Caldarelli, C.~Charmousis, and M.~Hassa\"\i{}ne, ``{AdS black holes with
  arbitrary scalar coupling},''
  \href{http://dx.doi.org/10.1007/JHEP10(2013)015}{{\em JHEP} {\bfseries 10}
  (2013) 015}, \href{http://arxiv.org/abs/1307.5063}{{\ttfamily arXiv:1307.5063
  [hep-th]}}.

\bibitem{BravoGaete:2013acu}
M.~Bravo~Gaete and M.~Hassaine, ``{Topological black holes for
  Einstein-Gauss-Bonnet gravity with a nonminimal scalar field},''
  \href{http://dx.doi.org/10.1103/PhysRevD.88.104011}{{\em Phys. Rev. D}
  {\bfseries 88} (2013) 104011},
  \href{http://arxiv.org/abs/1308.3076}{{\ttfamily arXiv:1308.3076 [hep-th]}}.

\bibitem{Erices:2024iah}
C.~Erices, L.~Guajardo, and K.~Lara, ``{Reverse stealth construction and its
  thermodynamic imprints},'' \href{http://arxiv.org/abs/2410.13719}{{\ttfamily
  arXiv:2410.13719 [gr-qc]}}.

\bibitem{Babichev:2013cya}
E.~Babichev and C.~Charmousis, ``{Dressing a black hole with a time-dependent
  Galileon},'' \href{http://dx.doi.org/10.1007/JHEP08(2014)106}{{\em JHEP}
  {\bfseries 08} (2014) 106}, \href{http://arxiv.org/abs/1312.3204}{{\ttfamily
  arXiv:1312.3204 [gr-qc]}}.

\bibitem{Kobayashi:2014eva}
T.~Kobayashi and N.~Tanahashi, ``{Exact black hole solutions in shift symmetric
  scalar\textendash{}tensor theories},''
  \href{http://dx.doi.org/10.1093/ptep/ptu096}{{\em PTEP} {\bfseries 2014}
  (2014) 073E02}, \href{http://arxiv.org/abs/1403.4364}{{\ttfamily
  arXiv:1403.4364 [gr-qc]}}.

\bibitem{Bravo-Gaete:2014haa}
M.~Bravo-Gaete and M.~Hassaine, ``{Thermodynamics of a BTZ black hole solution
  with an Horndeski source},''
  \href{http://dx.doi.org/10.1103/PhysRevD.90.024008}{{\em Phys. Rev. D}
  {\bfseries 90} no.~2, (2014) 024008},
  \href{http://arxiv.org/abs/1405.4935}{{\ttfamily arXiv:1405.4935 [hep-th]}}.

\bibitem{Minamitsuji:2018vuw}
M.~Minamitsuji and H.~Motohashi, ``{Stealth Schwarzschild solution in shift
  symmetry breaking theories},''
  \href{http://dx.doi.org/10.1103/PhysRevD.98.084027}{{\em Phys. Rev. D}
  {\bfseries 98} no.~8, (2018) 084027},
  \href{http://arxiv.org/abs/1809.06611}{{\ttfamily arXiv:1809.06611 [gr-qc]}}.

\bibitem{Minamitsuji:2019shy}
M.~Minamitsuji and J.~Edholm, ``{Black hole solutions in shift-symmetric
  degenerate higher-order scalar-tensor theories},''
  \href{http://dx.doi.org/10.1103/PhysRevD.100.044053}{{\em Phys. Rev. D}
  {\bfseries 100} no.~4, (2019) 044053},
  \href{http://arxiv.org/abs/1907.02072}{{\ttfamily arXiv:1907.02072 [gr-qc]}}.

\bibitem{Bernardo:2019yxp}
R.~C. Bernardo, J.~Celestial, and I.~Vega, ``{Stealth black holes in shift
  symmetric kinetic gravity braiding},''
  \href{http://dx.doi.org/10.1103/PhysRevD.101.024036}{{\em Phys. Rev. D}
  {\bfseries 101} no.~2, (2020) 024036},
  \href{http://arxiv.org/abs/1911.01847}{{\ttfamily arXiv:1911.01847 [gr-qc]}}.

\bibitem{Bakopoulos:2023fmv}
A.~Bakopoulos, C.~Charmousis, P.~Kanti, N.~Lecoeur, and T.~Nakas, ``{Black
  holes with primary scalar hair},''
  \href{http://dx.doi.org/10.1103/PhysRevD.109.024032}{{\em Phys. Rev. D}
  {\bfseries 109} no.~2, (2024) 024032},
  \href{http://arxiv.org/abs/2310.11919}{{\ttfamily arXiv:2310.11919 [gr-qc]}}.

\bibitem{Baake:2023zsq}
O.~Baake, A.~Cisterna, M.~Hassaine, and U.~Hernandez-Vera, ``{Endowing black
  holes with beyond-Horndeski primary hair: An exact solution framework for
  scalarizing in every dimension},''
  \href{http://dx.doi.org/10.1103/PhysRevD.109.064024}{{\em Phys. Rev. D}
  {\bfseries 109} no.~6, (2024) 064024},
  \href{http://arxiv.org/abs/2312.05207}{{\ttfamily arXiv:2312.05207
  [hep-th]}}.

\bibitem{Bakopoulos:2023sdm}
A.~Bakopoulos, N.~Chatzifotis, and T.~Nakas, ``{Compact objects with primary
  hair in shift and parity symmetric beyond Horndeski gravities},''
  \href{http://dx.doi.org/10.1103/PhysRevD.110.024044}{{\em Phys. Rev. D}
  {\bfseries 110} no.~2, (2024) 024044},
  \href{http://arxiv.org/abs/2312.17198}{{\ttfamily arXiv:2312.17198 [gr-qc]}}.

\bibitem{deRham:2019gha}
C.~de~Rham and J.~Zhang, ``{Perturbations of stealth black holes in degenerate
  higher-order scalar-tensor theories},''
  \href{http://dx.doi.org/10.1103/PhysRevD.100.124023}{{\em Phys. Rev. D}
  {\bfseries 100} no.~12, (2019) 124023},
  \href{http://arxiv.org/abs/1907.00699}{{\ttfamily arXiv:1907.00699
  [hep-th]}}.

\bibitem{Bernardo:2020ehy}
R.~C. Bernardo and I.~Vega, ``{Stealth black hole perturbations in kinetic
  gravity braiding},'' \href{http://dx.doi.org/10.1063/5.0048929}{{\em J. Math.
  Phys.} {\bfseries 62} no.~7, (2021) 072501},
  \href{http://arxiv.org/abs/2007.06006}{{\ttfamily arXiv:2007.06006 [gr-qc]}}.

\bibitem{Chagoya:2016aar}
J.~Chagoya, G.~Niz, and G.~Tasinato, ``{Black Holes and Abelian Symmetry
  Breaking},'' \href{http://dx.doi.org/10.1088/0264-9381/33/17/175007}{{\em
  Class. Quant. Grav.} {\bfseries 33} no.~17, (2016) 175007},
  \href{http://arxiv.org/abs/1602.08697}{{\ttfamily arXiv:1602.08697
  [hep-th]}}.

\bibitem{Heisenberg:2017xda}
L.~Heisenberg, R.~Kase, M.~Minamitsuji, and S.~Tsujikawa, ``{Hairy black-hole
  solutions in generalized Proca theories},''
  \href{http://dx.doi.org/10.1103/PhysRevD.96.084049}{{\em Phys. Rev. D}
  {\bfseries 96} no.~8, (2017) 084049},
  \href{http://arxiv.org/abs/1705.09662}{{\ttfamily arXiv:1705.09662 [gr-qc]}}.

\bibitem{Gripaios:2004ms}
B.~M. Gripaios, ``{Modified gravity via spontaneous symmetry breaking},''
  \href{http://dx.doi.org/10.1088/1126-6708/2004/10/069}{{\em JHEP} {\bfseries
  10} (2004) 069}, \href{http://arxiv.org/abs/hep-th/0408127}{{\ttfamily
  arXiv:hep-th/0408127}}.

\bibitem{Tasinato:2014eka}
G.~Tasinato, ``{Cosmic Acceleration from Abelian Symmetry Breaking},''
  \href{http://dx.doi.org/10.1007/JHEP04(2014)067}{{\em JHEP} {\bfseries 04}
  (2014) 067}, \href{http://arxiv.org/abs/1402.6450}{{\ttfamily arXiv:1402.6450
  [hep-th]}}.

\bibitem{Heisenberg:2017hwb}
L.~Heisenberg, R.~Kase, M.~Minamitsuji, and S.~Tsujikawa, ``{Black holes in
  vector-tensor theories},''
  \href{http://dx.doi.org/10.1088/1475-7516/2017/08/024}{{\em JCAP} {\bfseries
  08} (2017) 024}, \href{http://arxiv.org/abs/1706.05115}{{\ttfamily
  arXiv:1706.05115 [gr-qc]}}.

\bibitem{Chagoya:2017ojn}
J.~Chagoya and G.~Tasinato, ``{Stealth configurations in vector-tensor theories
  of gravity},'' \href{http://dx.doi.org/10.1088/1475-7516/2018/01/046}{{\em
  JCAP} {\bfseries 01} (2018) 046},
  \href{http://arxiv.org/abs/1707.07951}{{\ttfamily arXiv:1707.07951
  [hep-th]}}.

\bibitem{Minamitsuji:2016ydr}
M.~Minamitsuji, ``{Solutions in the generalized Proca theory with the
  nonminimal coupling to the Einstein tensor},''
  \href{http://dx.doi.org/10.1103/PhysRevD.94.084039}{{\em Phys. Rev. D}
  {\bfseries 94} no.~8, (2016) 084039},
  \href{http://arxiv.org/abs/1607.06278}{{\ttfamily arXiv:1607.06278 [gr-qc]}}.

\bibitem{Ayon-Beato:2013bsa}
E.~Ay\'on-Beato, A.~A. Garc\'\i{}a, P.~I. Ram\'\i{}rez-Baca, and C.~A.
  Terrero-Escalante, ``{Conformal stealth for any standard cosmology},''
  \href{http://dx.doi.org/10.1103/PhysRevD.88.063523}{{\em Phys. Rev. D}
  {\bfseries 88} no.~6, (2013) 063523},
  \href{http://arxiv.org/abs/1307.6534}{{\ttfamily arXiv:1307.6534 [gr-qc]}}.

\bibitem{Ayon-Beato:2015mxf}
E.~Ay\'on-Beato, P.~I. Ram\'\i{}rez-Baca, and C.~A. Terrero-Escalante,
  ``{Cosmological stealths with nonconformal couplings},''
  \href{http://dx.doi.org/10.1103/PhysRevD.97.043505}{{\em Phys. Rev. D}
  {\bfseries 97} no.~4, (2018) 043505},
  \href{http://arxiv.org/abs/1512.09375}{{\ttfamily arXiv:1512.09375 [gr-qc]}}.

\bibitem{Maeda:2012tu}
H.~Maeda and K.-i. Maeda, ``{Creation of the universe with a stealth scalar
  field},'' \href{http://dx.doi.org/10.1103/PhysRevD.86.124045}{{\em Phys. Rev.
  D} {\bfseries 86} (2012) 124045},
  \href{http://arxiv.org/abs/1208.5777}{{\ttfamily arXiv:1208.5777 [gr-qc]}}.

\bibitem{Bakopoulos:2024hah}
A.~Bakopoulos, T.~Karakasis, N.~E. Mavromatos, T.~Nakas, and
  E.~Papantonopoulos, ``{Exact black holes in string-inspired Euler-Heisenberg
  theory},'' \href{http://dx.doi.org/10.1103/PhysRevD.110.024014}{{\em Phys.
  Rev. D} {\bfseries 110} no.~2, (2024) 024014},
  \href{http://arxiv.org/abs/2402.12459}{{\ttfamily arXiv:2402.12459
  [hep-th]}}.

\bibitem{Karakasis:2023hni}
T.~Karakasis, N.~E. Mavromatos, and E.~Papantonopoulos, ``{Regular compact
  objects with scalar hair},''
  \href{http://dx.doi.org/10.1103/PhysRevD.108.024001}{{\em Phys. Rev. D}
  {\bfseries 108} no.~2, (2023) 024001},
  \href{http://arxiv.org/abs/2305.00058}{{\ttfamily arXiv:2305.00058 [gr-qc]}}.

\bibitem{Gibbons:1976ue}
G.~W. Gibbons and S.~W. Hawking, ``{Action Integrals and Partition Functions in
  Quantum Gravity},'' \href{http://dx.doi.org/10.1103/PhysRevD.15.2752}{{\em
  Phys. Rev. D} {\bfseries 15} (1977) 2752--2756}.

\bibitem{Bakopoulos:2024ogt}
A.~Bakopoulos, N.~Chatzifotis, and T.~Karakasis, ``{Thermodynamics of black
  holes featuring primary scalar hair},''
  \href{http://dx.doi.org/10.1103/PhysRevD.110.L101502}{{\em Phys. Rev. D}
  {\bfseries 110} no.~10, (2024) L101502},
  \href{http://arxiv.org/abs/2404.07522}{{\ttfamily arXiv:2404.07522
  [hep-th]}}.

\bibitem{Bakopoulos:2023tso}
A.~Bakopoulos, N.~Chatzifotis, C.~Erices, and E.~Papantonopoulos, ``{Stealth
  Ellis wormholes in Horndeski theories},''
  \href{http://dx.doi.org/10.1088/1475-7516/2023/11/055}{{\em JCAP} {\bfseries
  11} (2023) 055}, \href{http://arxiv.org/abs/2306.16768}{{\ttfamily
  arXiv:2306.16768 [hep-th]}}.

\bibitem{Kobayashi:2011nu}
T.~Kobayashi, M.~Yamaguchi, and J.~Yokoyama, ``{Generalized G-inflation:
  Inflation with the most general second-order field equations},''
  \href{http://dx.doi.org/10.1143/PTP.126.511}{{\em Prog. Theor. Phys.}
  {\bfseries 126} (2011) 511--529},
  \href{http://arxiv.org/abs/1105.5723}{{\ttfamily arXiv:1105.5723 [hep-th]}}.

\bibitem{Minamitsuji:2023nvh}
M.~Minamitsuji and K.-i. Maeda, ``{Black hole thermodynamics in Horndeski
  theories},'' \href{http://dx.doi.org/10.1103/PhysRevD.108.084061}{{\em Phys.
  Rev. D} {\bfseries 108} no.~8, (2023) 084061},
  \href{http://arxiv.org/abs/2308.01082}{{\ttfamily arXiv:2308.01082 [gr-qc]}}.

\bibitem{Wald:1993nt}
R.~M. Wald, ``{Black hole entropy is the Noether charge},''
  \href{http://dx.doi.org/10.1103/PhysRevD.48.R3427}{{\em Phys. Rev. D}
  {\bfseries 48} no.~8, (1993) R3427--R3431},
  \href{http://arxiv.org/abs/gr-qc/9307038}{{\ttfamily arXiv:gr-qc/9307038}}.

\bibitem{Jacobson:1993vj}
T.~Jacobson, G.~Kang, and R.~C. Myers, ``{On black hole entropy},''
  \href{http://dx.doi.org/10.1103/PhysRevD.49.6587}{{\em Phys. Rev. D}
  {\bfseries 49} (1994) 6587--6598},
  \href{http://arxiv.org/abs/gr-qc/9312023}{{\ttfamily arXiv:gr-qc/9312023}}.

\bibitem{Hajian:2020dcq}
K.~Hajian, S.~Liberati, M.~M. Sheikh-Jabbari, and M.~H. Vahidinia, ``{On Black
  Hole Temperature in Horndeski Gravity},''
  \href{http://dx.doi.org/10.1016/j.physletb.2020.136002}{{\em Phys. Lett. B}
  {\bfseries 812} (2021) 136002},
  \href{http://arxiv.org/abs/2005.12985}{{\ttfamily arXiv:2005.12985 [gr-qc]}}.

\bibitem{Brustein:2007jj}
R.~Brustein, D.~Gorbonos, and M.~Hadad, ``{Wald's entropy is equal to a quarter
  of the horizon area in units of the effective gravitational coupling},''
  \href{http://dx.doi.org/10.1103/PhysRevD.79.044025}{{\em Phys. Rev. D}
  {\bfseries 79} (2009) 044025},
  \href{http://arxiv.org/abs/0712.3206}{{\ttfamily arXiv:0712.3206 [hep-th]}}.

\bibitem{Iyer:1994ys}
V.~Iyer and R.~M. Wald, ``{Some properties of Noether charge and a proposal for
  dynamical black hole entropy},''
  \href{http://dx.doi.org/10.1103/PhysRevD.50.846}{{\em Phys. Rev. D}
  {\bfseries 50} (1994) 846--864},
  \href{http://arxiv.org/abs/gr-qc/9403028}{{\ttfamily arXiv:gr-qc/9403028}}.

\bibitem{PhysRevD.74.044007}
S.~Dutta and R.~Gopakumar, ``Euclidean and noetherian entropies in ads space,''
  \href{http://dx.doi.org/10.1103/PhysRevD.74.044007}{{\em Phys. Rev. D}
  {\bfseries 74} (Aug, 2006) 044007}.
  \url{https://link.aps.org/doi/10.1103/PhysRevD.74.044007}.

\bibitem{Tachikawa:2006sz}
Y.~Tachikawa, ``{Black hole entropy in the presence of Chern-Simons terms},''
  \href{http://dx.doi.org/10.1088/0264-9381/24/3/014}{{\em Class. Quant. Grav.}
  {\bfseries 24} (2007) 737--744},
  \href{http://arxiv.org/abs/hep-th/0611141}{{\ttfamily arXiv:hep-th/0611141}}.

\bibitem{Bekenstein:1973ur}
J.~D. Bekenstein, ``{Black holes and entropy},''
  \href{http://dx.doi.org/10.1103/PhysRevD.7.2333}{{\em Phys. Rev. D}
  {\bfseries 7} (1973) 2333--2346}.

\bibitem{Hawking:1975vcx}
S.~W. Hawking, ``{Particle Creation by Black Holes},''
  \href{http://dx.doi.org/10.1007/BF02345020}{{\em Commun. Math. Phys.}
  {\bfseries 43} (1975) 199--220}. [Erratum: Commun.Math.Phys. 46, 206 (1976)].

\bibitem{Feng:2015oea}
X.-H. Feng, H.-S. Liu, H.~L\"u, and C.~N. Pope, ``{Black Hole Entropy and
  Viscosity Bound in Horndeski Gravity},''
  \href{http://dx.doi.org/10.1007/JHEP11(2015)176}{{\em JHEP} {\bfseries 11}
  (2015) 176}, \href{http://arxiv.org/abs/1509.07142}{{\ttfamily
  arXiv:1509.07142 [hep-th]}}.

\end{thebibliography}\endgroup
\bibliographystyle{utphys}

\end{document}